\documentclass[journal=jacsat,manuscript=communication,articletitle=true]{achemso}
\usepackage[version=3]{mhchem} % Formula subscripts using \ce{}
\DeclareUnicodeCharacter{2212}{\textminus}
\usepackage{booktabs}
\usepackage{makecell} 
\usepackage{graphicx}
\usepackage[T1]{fontenc}       % Use modern font encodings
\usepackage{dblfloatfix}
\usepackage[inline]{enumitem}
\usepackage{gensymb}
\usepackage{grffile}
\usepackage{chngcntr}
\usepackage{xcolor}
\usepackage[dvipsnames]{xcolor}
\usepackage{chemformula}
\usepackage[T1]{fontenc} % Use modern font encodings
\usepackage{epsfig}
\usepackage{braket}
\usepackage{psfrag}
\usepackage{subfig}
\usepackage{multirow}
\usepackage{xr}
\usepackage{mathtools}
\usepackage{amsmath}
\usepackage{amssymb}
\usepackage{commath}
\usepackage{amsfonts}
\usepackage[normalem]{ulem}
\usepackage{braket}
\usepackage{array}
\usepackage{tabularx}
\usepackage{stackengine}
\usepackage{hyperref}
\usepackage[utf8]{inputenc}
\usepackage{newunicodechar}
\newunicodechar{₂}{$_2$}
\newunicodechar{₄}{$_4$}
\usepackage{fix-cm}
\usepackage[table]{xcolor}
\definecolor{lightblue}{RGB}{200,220,255}
\definecolor{lightgreen}{RGB}{200,255,200}
\definecolor{lightyellow}{RGB}{255,255,180}
\usepackage{algorithm}
\usepackage{algpseudocode}
\usepackage{textcomp}

\SectionsOn

\setkeys{acs}{articletitle = true}
\author{Ashna Jose}
\affiliation{Univ. Grenoble Alpes, CNRS, Grenoble INP, SIMaP, 38000 Grenoble, France}
\altaffiliation{Department of Materials, Imperial College London, London SW7 2AZ, United Kingdom}
\email{a.jose@imperial.ac.uk}

\author{Emilie Devijver}
\affiliation[]
{Univ. Grenoble Alpes, CNRS, Grenoble INP, LiG, 38000 Grenoble, France}

\author{Martin Uhrin}
\affiliation[]
{Univ. Grenoble Alpes, CNRS, Grenoble INP, SIMaP, 38000 Grenoble, France}

\author{Noel Jakse}
\affiliation[]
{Univ. Grenoble Alpes, CNRS, Grenoble INP, SIMaP, 38000 Grenoble, France}

\author{Roberta Poloni}
\affiliation[]
{Univ. Grenoble Alpes, CNRS, Grenoble INP, SIMaP, 38000 Grenoble, France}
\email{roberta.poloni@grenoble-inp.fr}

\title[An \textsf{achemso} demo]
  {Predicting Spin-Crossover Behavior in Metal-Organic Frameworks from Limited and Noisy Data Using Quantile Active Learning}

\abbreviations{}

\begin{document}

%\begin{tocentry}
%\includegraphics{jacs_graphical_abstract.pdf}
%\end{tocentry}

\maketitle
\begin{abstract}
Spin-crossover (SCO) metal-organic frameworks (MOFs) hold great promise for sensing, spintronics, and gas-related applications, however, only a small number of SCO-active examples are known among the thousands of MOFs already synthesized. Computational screening enhanced by machine learning offers a powerful route to uncover these hidden candidates much more rapidly than trial-and-error experiments. However, progress is limited by the computational complexity of obtaining accurate adiabatic energy differences, as these typically require separate geometry optimizations for both spin states, a process that is technically challenging, prone to convergence failures, and difficult to automate at scale. To mitigate these issues, we introduce a data-efficient strategy based on Quantile Regression Tree-based Active Learning, designed to navigate large chemical spaces while remaining robust to noisy and scarce labels obtained from unrelaxed geometries.
After actively selecting a 200-sized subset of representative MOFs for electronic-structure evaluation, a Random Forest regressor trained on this data accurately identifies SCO-relevant candidates despite label noise, recovering $\sim$82\% of true positives with only two false negatives.
Applying the model to the unlabeled dataset yields a new collection of high-confidence SCO MOFs, which we denote pSCO-105. This work shows that spin crossover can be reliably identified from limited and imperfect data through smart training-set selection, enabling accelerated screening of SCO MOFs.

\end{abstract}

\section{Introduction}

Spin-crossover (SCO) complexes and materials \cite{Gutlich2006, Gutlich2006-tk, Gutlich2004-ha, Takahashi2018} featuring transition-metal centers capable of switching between low-spin (LS) and high-spin (HS) electronic configurations, have long attracted attention due to their potential applications in molecular spintronics \cite{D1NA00407G}, memory devices \cite{B306638J}, sensors \cite{Linares2012-rt}, optical devices \cite{Gutlich2004-ha}, and gas capture \cite{PMID:28892810, AMariano2023}. 
Metal-organic frameworks (MOFs \cite{Zhou2012-wv, doi:10.1021/ja308229p}) represent an attractive platform for SCO, combining structural tunability, porosity, and functionality, possibly enabling the development of switchable adsorbents for gas-separation materials as suggested \cite{Poloni2014} and demonstrated by a few authors \cite{Reed2016-cq, AMariano2023}. While thousands of MOFs have already been reported in experimental databases\cite{qmof2,core} and are synthesisable through established protocols, only a few SCO-active MOFs have been identified and studied to date. Before engaging in \textit{de novo} design efforts, it is therefore essential to evaluate whether SCO behavior may already be present in these existing materials using advanced computational screening techniques.
Data-driven approaches such as machine learning (ML) and particularly active learning (AL) have emerged as powerful tools for accelerating functional MOF discovery \cite{jacsau.4c00618} by predicting properties and guiding the selection of informative candidates for a wide range of applications \cite{doi:10.1021/acs.chemrev.0c00004, Jablonka2024, Elena2025-rw, Gibaldi2025, Lim2025, Mourino2025} such as gas capture \cite{SEYYEDATTAR2024158, Wang2024-qh, JERNG2024100361, doi:10.1021/acs.jpcc.6b08729, Boyd2019}, prediction of heat capacities \cite{Moosavi2022}, electronic band gaps \cite{ROSEN20211578, qmof2}, and partial atomic charges \cite{arcmof}. Various ML algorithms such as tree-based models \cite{jacs_ashna, SEYYEDATTAR2024158, racs}, Gaussian processes \cite{nandy, gp_mof1, gp_mof3, gp_mof4}, artificial neural networks \cite{racs, woo}, and transformer models \cite{moformer, moftransformer, pmtransformer} have been used to accelerate MOF discovery and understand structure-property relationships. 
Recent works have applied ML to predict spin-transition energetics \cite{Janet2017-mv, doi:10.1021/acs.jpca.3c07104}. However, these studies largely focus on Fe(II)-based transition-metal complexes rather than MOFs. 
The quantity relevant to establish the feasibility of SCO is the adiabatic energy difference, $\Delta E_\text{H--L} = E_{HS} - E_{LS}$, between the high spin (HS) and low-spin (LS) states.
Notably, accurately predicting this quantity in large systems that exhibit a large structural diversity, such as MOFs, is notoriously difficult. This is because accurate density functional theory (DFT) is computationally expensive and challenging \cite{MEJIARODRIGUEZ2022111161, Mariano2020-nx, Mariano2021-wu, trular2020,KEPP2013196, Vennelakanti2023-rc}, and because obtaining high-quality reference labels requires separate geometry optimizations of the HS and LS states. In practice, these optimizations frequently suffer from convergence difficulties, broken symmetries, or require substantial manual intervention, making high-throughput screening impractical for large MOF datasets. 

\begin{figure*}[t]
\centering
\includegraphics[scale=0.53,
trim=0.5cm 3.9cm 0cm 1.8cm,
clip]{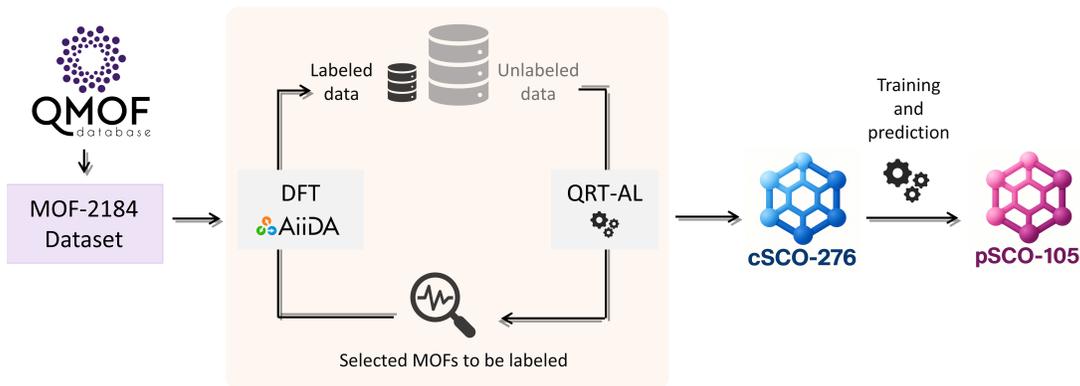}
\caption{\textbf{Schematic representation of the workflow developed in this work.} The QMOF database is pre-screened for potential SCO-active MOFs, and the MOF-2184 subset is obtained. The test set is selected using a clustering-based algorithm. QRT-AL is then initialized with a randomly selected subset of MOFs, for which $\Delta E_\text{H--L}$ values are computed using DFT (AiiDA). Labeled MOFs are added to the training set, and the active learning loop is iterated until 200 MOFs are labeled. The resulting set of 276 $\Delta E_\text{H--L}$ values computed using DFT without geometrical optimization is named here cSCO-276 dataset. This set
is used to train a machine learning model that predicts $\Delta E_\text{H--L}$, using which high-confidence SCO-active MOFs (pSCO-105) are identified.}
\label{fig:1}
\end{figure*}

To mitigate these limitations, we propose leveraging Quantile Regression Tree-based Active Learning (QRT-AL) to efficiently sample training data from a database consisting of $\Delta E_\text{H--L}$ values computed for unrelaxed structures. While this approximation greatly increases screening speed, it also introduces label noise, since, in reality, LS and HS geometries may differ appreciably from one another leading to inaccuracies in our calculated $\Delta E_\text{H--L}$. Building on our earlier Regression Tree-based Active Learning (RT-AL) framework \cite{jacs_ashna}, which identifies diverse and informative samples across a chemical space, the QRT-AL strategy introduced recently \cite{qrtal} incorporates quantile estimates to focus sampling towards materials with specific properties, ensuring good predictions on a quantile of interest. Specifically, we address the prediction of MOFs that exhibit $\Delta E_\text{H--L}$ values compatible with spin-crossover behavior.
The general methodology developed in this work is outlined schematically in Figure \ref{fig:1}.
We establish a complete workflow to accelerate the discovery of SCO-active MOFs using limited electronic-structure calculations. Starting from the QMOF database \cite{qmof2}, we curate a focused MOF-2184 dataset containing MOFs with a unique transition metal, whose electron count is consistent with a spin crossover. Using QRT-AL, we iteratively select MOFs for $\Delta E_\text{H--L}$ labeling through automated DFT calculations conducted via a custom AiiDA workflow \cite{aiida, martin_uhrin} developed in this work. Once a sufficiently informative training set is acquired, a Random Forest regression model is trained to predict energy differences across the curated dataset, enabling rapid identification of promising SCO candidates.

Despite a moderate mean absolute error on the unseen test set, the model performs well when framed as a binary classifier for identifying SCO-relevant MOFs. It achieves a recall of 82\% with only two false negatives among the highly curated test set for which $\Delta E_\text{H--L}$ values optimized at the respective spin states were computed.
%Despite a moderate mean absolute error on an unseen test set, when evaluated as a binary classifier distinguishing SCO-relevant $\Delta E_\text{H--L}$ values, the model achieves a recall of 82\%, with only two false negatives among the highly curated test set for which $\Delta E_\text{H--L}$ values optimized at the respective spin states were computed. 
This confirms its robustness in capturing true SCO candidates while keeping the risk of discarding valuable materials extremely low. Our model also correctly identifies out-of-distribution SCO-active molecules, complexes, and MOFs previously reported in the literature.
Finally, applying the trained model across the unlabeled part of the dataset, we uncover a set of 105 MOFs whose predicted $\Delta E_\text{H--L}$ values lie within the target SCO window and are associated with high confidence. This computationally curated collection is referred to as the pSCO-105 set. 
This work establishes that promising SCO-active MOFs can be identified reliably and with high confidence, even from noisy data. Furthermore, we show that integration of targeted active learning with automated electronic-structure workflows provides a powerful and scalable framework for exploring vast chemical spaces and accelerating the discovery of complex spin-crossover materials.
%This work demonstrates that even in the presence of noisy labels, promising SCO-active MOFs can be efficiently identified with a high level of confidence.
Our study shows that combining targeted active learning with automated electronic-structure workflows enables efficient navigation of large and diverse chemical spaces to identify complex spin-crossover behavior.

\section{Results}

\begin{figure*}[ht]
\centering
\includegraphics[scale=0.57,
trim=0cm 0.5cm 1cm 1cm,
clip]{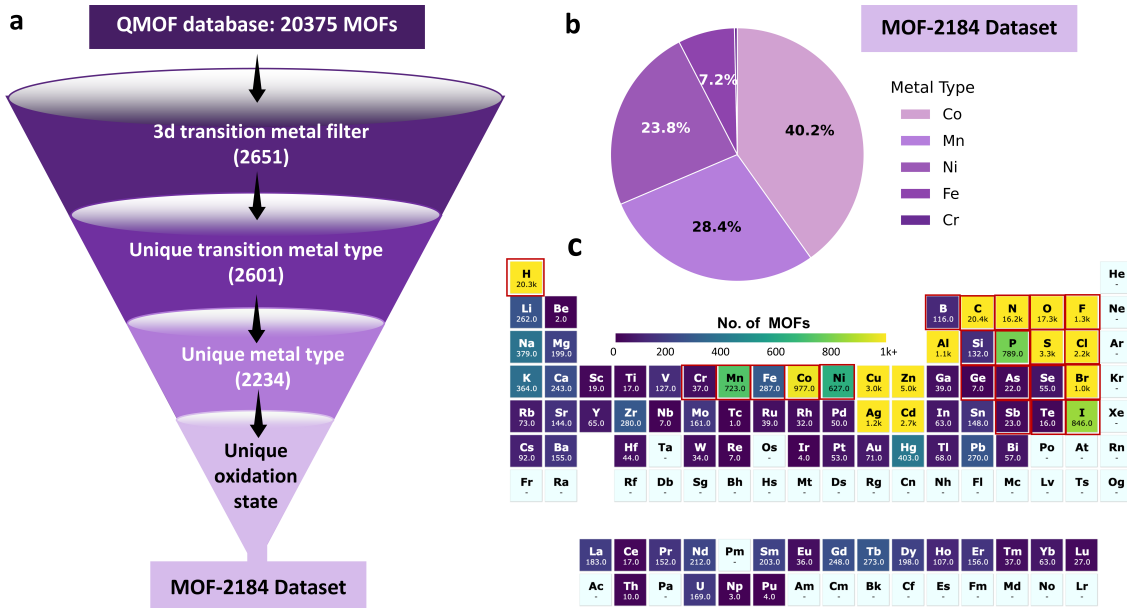}
\caption{(a) Overview of the pre-screening steps applied to the QMOF database to obtain the MOF-2184 subset, containing potential SCO candidates.
(b) Pie chart showing the distribution of MOFs with various transition metals.
(c) Periodic table showing the number of MOFs with a given element in the QMOF database, the elements that are present in the MOF-2184 subset are highlighted in red.}
\label{fig:2}
\end{figure*}

\subsection{MOF-2184 Dataset}
The QMOF database \cite{qmof2}, comprising 20,375 MOFs, served as the starting point for this work to curate a subset of potential SCO candidates. 
Only structures containing at least one first-row transition metal (i.e. Cr, Mn, Fe, Co and Ni) identified using Atomic Simulation Environment (ASE \cite{ase-paper, ISI:000175131400009}) were selected, consistent with the spin crossover phenomenon. This resulted in a subset of 2651 MOFs. To simplify the workflow and enable automation, only MOFs containing a single type of transition metal were considered, more strictly, those with only one type of metal overall, resulting in a subset of 2234 MOFs.
%To perform spin polarized \textit{ab initio} calculations at both HS and LS states, the total magnetization of these states for each MOF was determined.
Oxidation states of the transition metals were computed using oxiMACHINE \cite{Jablonka2021}. By considering the local environment around the metal, oxiMACHINE employs classification-based machine learning models to predict oxidation states in MOFs. Using these predictions, MOFs containing transition metals with multiple possible oxidation states were excluded, resulting in the final subset of 2184 structures, hereafter referred to as the MOF-2184 dataset. A summary of all pre-screening steps is provided in Figure \ref{fig:2}(a). 

Figure \ref{fig:2}(b) shows the distribution of metal atom types in the final dataset, with MOFs containing Co being most abundant (40.2\%) and MOFs containing Cr the least (0.4\%). The elements present in the MOF-2184 dataset are highlighted in the periodic table presented in Figure \ref{fig:2}(c). In addition, all possible electronic configurations represented in the dataset are summarized in Table \ref{tab:oximachine}, along with their LS and HS spin multiplicities. Note that while some cases may also exhibit intermediate spin states, they are not considered in the context of this work, and we define $\Delta E_\text{H--L}$ to be the energy difference between the lowest and highest spin configurations, even if there the energy difference to an intermediate spin state is lower. Finally, the total magnetization, as determined by the oxidation state and spin, is fixed during the DFT calculations (see the Materials and Methods section).

\begin{table}[t]
\caption{Spin multiplicities for low-spin (LS) and high-spin (HS) states of 3d transition-metal electronic configurations present in the MOF-2184 dataset. The corresponding total magnetization values, computed from the oxidation states, are also reported.}
\label{tab:oximachine}
\centering
\renewcommand{\arraystretch}{1.1}
\begin{tabular*}{\columnwidth}{@{\extracolsep{\fill}}lcccc}
\toprule
Configuration & \multicolumn{2}{c}{Multiplicity} & \multicolumn{2}{c}{Magnetization} \\
             & LS & HS & LS & HS \\

\midrule
d$^3$ (Cr$^{3+}$)                 & 1/2 & 3/2 & 1 & 3 \\
d$^4$ (Cr$^{2+}$, Mn$^{3+}$)      & 0   & 2   & 0 & 4 \\
d$^5$ (Fe$^{3+}$, Mn$^{2+}$)      & 1/2 & 5/2 & 1 & 5 \\
d$^6$ (Co$^{3+}$, Fe$^{2+}$)      & 0   & 2   & 0 & 4 \\
d$^7$ (Co$^{2+}$)                 & 1/2 & 3/2 & 1 & 3 \\
d$^8$ (Ni$^{2+}$)                 & 0   & 1   & 0 & 2 \\
\bottomrule
\end{tabular*}
\end{table}

\subsection{Test set construction}

\begin{figure}[h]
\centering
\vspace{0.2cm}
\includegraphics[scale=0.42]{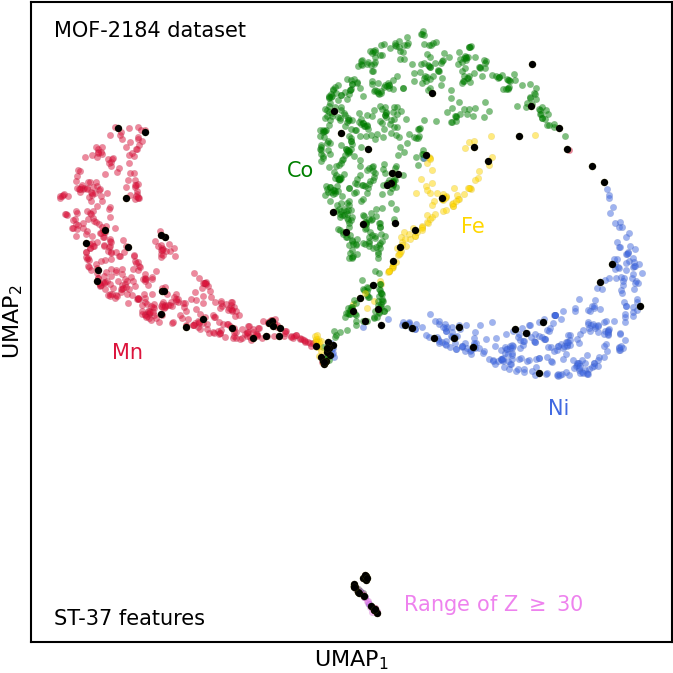} \vspace{0.1cm}
\caption{UMAP visualization of the MOF-2184 dataset in the ST-37 descriptor space. Five prominent clusters can be observed: four major clusters correspond to MOFs containing transition metals Fe (yellow), Co (green), Ni (blue), and Mn (red). A smaller cluster at the bottom (violet) consists of MOFs with the \textit{range of atomic number} feature $\geq$ 30. Black circles indicate MOFs selected by iRDM in the test set, showing a well-distributed selection across the different clusters.}
\label{fig:3}
\end{figure}

Because the MOF-2184 dataset lacks predefined labels, no \textit{a priori} test set is available to assess the performance of trained machine-learning models. To ensure a representative and informative evaluation subset, we applied the representativity-based sampling method Iterative Representativeness Diversity Maximization (iRDM \cite{irdm}) to select 100 MOFs, instead of randomly choosing them. iRDM partitions the feature space of the given data into clusters, and samples are selected such that they are diverse and representative of these clusters. 
The features used to select MOFs via iRDM were the 37 non-zero features of the Stoichiometric-120 feature-set (ST-37 \cite{st120}), which consists of element fractions and statistical attributes such as mean and range of atomic number, atomic mass, electronegativities etc (see S1 for more details). The two-dimensional UMAP \cite{umap} representation of the MOF-2184 dataset in the ST-37 feature space, presented in Figure \ref{fig:3}, shows that this descriptor partitions the dataset into clusters characterized by the transition metal and ligand chemistry. This underlies our choice of descriptor here: samples that are representative in this descriptor space ensure a test set that captures diversity in metal and ligand chemistry, which is crucial for fairly testing the trained ML models. As illustrated by the UMAP, the test set (shown as black circles) is indeed diverse and representative. Note, however, that since we do not yet have access to the label, using iRDM for selecting the training set is not ideal, as it does not guarantee representativeness with respect to the unknown label distribution.

Subsequently, the $\Delta E_\text{H--L}$ values were computed for the selected test set using a custom AiiDA Quantum ESPRESSO workflow developed in this work. Initially the \texttt{SCO-MOF-RelaxWorkChain} (described in detail in Methods and S2) was used to compute $\Delta E_\text{H--L}$. In this scheme, a geometry optimization was performed for each MOF at each spin state, by setting a fixed total magnetization. 
However, the workflow achieved convergence for both spin states for only 50 of the 100 MOFs.
Difficulties in achieving convergence are related to the high energy of the metastable spin state, but also due to self-consistent field (SCF) instabilities and numerical issues that are common in transition-metal systems, despite allowing a large number of iterations.
By contrast employing the \texttt{SCO-MOF-SCF-WorkChain}, which uses the relaxed structure previously optimized at the PBE-D3(BJ) level in the work of Rosen et al. \cite{qmof2}, but not optimized for both spin states, was significantly more successful. In this case, convergence was achieved for the vast majority of structures, with only 13 MOFs failing to converge with the parameters described in the Methods section. Since data-driven approaches rely heavily on the availability of sufficient data, the \texttt{SCO-MOF-SCF-WorkChain} was ultimately adopted for all calculations in this work, and $\Delta E_\text{H--L}$ values were computed using the ground-state geometry. 

Note that although geometry optimization is not performed at each spin state for the final set of labels, these values still provide a reasonable estimate and serve as a foundation for studying this complex phenomenon through a combination of ab initio methods and machine learning. All the labels computed hereafter are obtained using the \texttt{SCO-MOF-SCF-WorkChain}, where the structure is taken from the MOF-2184 database and the SCF calculations are performed for the LS and HS states, with fixed total magnetization.

\subsection{Quantile Active Learning}

Because evaluating both high- and low-spin states is computationally demanding, active learning provides an efficient data-driven strategy by iteratively selecting the most informative and diverse samples, thereby minimizing redundant and costly labeling. Many active learning methods have been proposed in the past \cite{gsx, irdm, cube, qbc, emcm, deepl, gp, mondrianTrees, ashna} and have been exploited in materials informatics \cite{gp_mof1, gp_mof4, jacs_ashna} as well. 
Here, beyond ensuring informative sampling alone, a strategy that actively targets a range of particularly relevant values is more appropriate, since the goal of this work is to accurately predict $\Delta E_\text{H--L}$ values within the SCO regime. To that end, we employ an extension of our Regression Tree-based Active Learning (RT-AL \cite{ashna}) approach. RT-AL leverages both input features and output trends to iteratively construct an optimal training set. The extension used here is Quantile RT-AL (QRT-AL \cite{qrtal}), which is specifically designed for situations where a given quantile of the output distribution is of primary interest. Tests on MOF datasets have shown \cite{qrtal} that targeting a specific quantile enables efficient learning of band gaps and gas adsorption with very limited labeled data.

The method operates as follows: an initial training set, $I_{\text{init}}$, of size $n_{\text{init}}$ is constructed either randomly, or using an input-feature based AL method \cite{gsx, irdm, cube}, from a pool of unlabeled data which is then labeled. A regression tree with $K$ leaves is then trained, and is used to predict labels of the remaining unlabeled samples. The leaves of this tree are thereafter used to add more samples to the training set. The number of samples to be labeled from each leaf $k$, $n_k^*$, are distributed into different leaves as shown below:

\begin{equation*}
n_k^* = n_{\text{act}} \frac{\sqrt{\pi_k \hat \sigma_k^2 \gamma_k}}{\sum_{\ell=1}^K \sqrt{\pi_\ell \hat \sigma_\ell^2 \gamma_l}}.
\end{equation*}

Here $n_{\text{act}}$ is the total number of samples to be selected by QRT-AL in the current iteration, $\hat \sigma_k^2$ denotes the variance computed from the true values of the labels in leaf $k$, and $\pi_k$ is the proportion of unlabeled samples in leaf $k$. $\gamma_k$ specifies the quantile interval of interest, such that for each leaf $1 \leq k \leq K$,

\begin{equation*}
 \gamma_k = \frac{\sum_{q=1}^Q {w^q n^q_k }}{\sum_{q=1}^Q n^q_k}.
 \end{equation*}

\begin{figure}[t]
\centering
\includegraphics[scale=0.56,
trim=7.5cm 0cm 6.5cm 0cm,
clip]{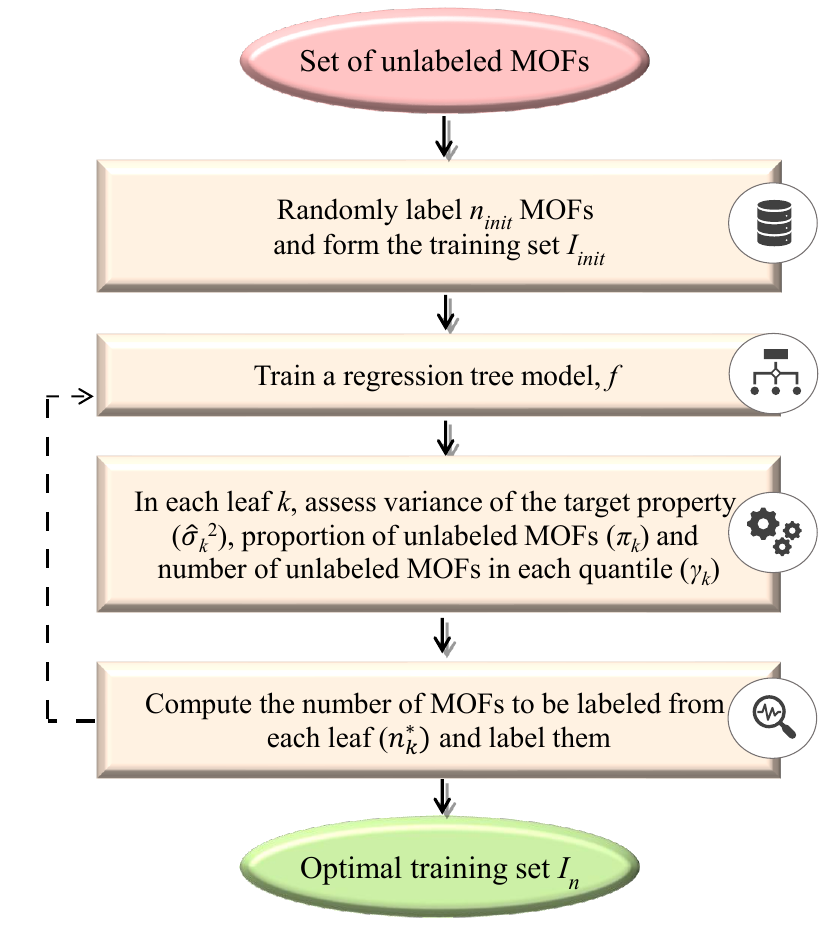}
\caption{Overview of training set construction using Quantile Regression Tree–Based Active Learning (QRT-AL)}
\label{fig:4}
\end{figure}

In the above equation, $n^q_k$ are the number of unlabeled samples in leaf $k$ in quantile interval $q$, and $w^q$ are weights defined depending on the quantile of interest. Thus, the number of samples to be labeled depends on:
\begin{enumerate*}[label=(\roman*)]
    \item variance computed from the true labels,
    \item the probability that an unlabeled sample belongs to a given leaf, and
    \item the range of the target values in each leaf, thereby sampling more from regions that belong to the quantile of interest.
\end{enumerate*}
The quantiles and weights are chosen based on the region of interest: the set primarily targets this region while retaining a small number of samples from other quantiles to maintain a global view of the dataset and reduce overfitting. Accordingly, the range of known target values is divided into $Q$ quantile intervals. Details of the quantiles and their associated weights are given in the following section. The QRT-AL algorithm is presented below and is summarized in Figure \ref{fig:4}.

\begin{algorithm}
\caption{Quantile Regression Tree-based Active Learning (QRT-AL)}\label{alg_qrtal}
\textbf{Input}: Labeled set: $(\mathbf x_i, y_i)_{i \in I_{\text{init}}}$, unlabeled set: $(\mathbf x_i)_{i\notin I_{\text{init}}}$; $n_{\text{act}}$: maximum new samples; quantile interval: $Q$  

\begin{algorithmic}[1]
\State Construct a standard regression tree with $K$ leaves using $(\mathbf x_i, y_i)_{i \in I_{\text{init}}}$

\For{$k = 1,\ldots,K$}
  \State Compute $\pi_k$, $\hat \sigma_k^2$, and $\gamma_k$
  \State Calculate the number of samples $n_k^*$ to be labeled from leaf $k$
  \State Identify $I_{\text{act}}^k$, the set of $n_k^*$ observations from leaf $k$
\EndFor

\end{algorithmic}

{\raggedright
\textbf{Output}: The set $\cup_{k=1}^K (\mathbf x_i)_{i \in I_{\text{act}}^k}$ of observations to be labeled \par}

\end{algorithm}

The samples added to the training set via QRT-AL are diverse and representative in both input and target spaces, while specifically focusing on the quantile of interest. This targeted sampling enables more effective learning within the relevant region, as demonstrated in Ref.~\cite{qrtal}.
After computing the number of samples to be selected from each region, these are selected using random sampling within the same. Then, the regression tree is retrained, leading to a more accurate model. The active learning cycle is repeated, and new samples are labeled and added to the training set at each step, followed by retraining the regression tree, until the desired training set size is achieved, or a targeted accuracy of the model is reached. In this work, up to 200 MOFs are selected in the training set using this scheme. Finally, ensemble tree-based models \cite{rf} are trained, and their performance is evaluated using the test set selected earlier.

\subsection{Parameters for QR-TAL}
\paragraph{Adiabatic energy window for SCO} For selecting the training set, we consider $\Delta E_\text{H--L}$ values between 0 and 1 eV as relevant for a near-room-temperature SCO. This is justified as follows. The spin-crossover transition temperature, $T_{1/2}$, is obtained from the equilibrium condition
$\Delta G = \Delta H - T_{1/2}\Delta S = 0$.
The enthalpy difference, $\Delta H$, represents the adiabatic electronic energy difference between the high- and low-spin states, together with zero-point energy, and finite-temperature vibrational contributions.
Assuming entropic contributions in the range $\sim 9$--$90$~J~mol$^{-1}$~K$^{-1}$, as reported for a broad class of transition-metal complexes,\cite{Kepp2016,Vela2020}
a room-temperature spin crossover ($T_{1/2}\approx 300$~K) corresponds to target values of $\Delta H$ between approximately $0$ and $0.3$~eV.
Previous studies have shown that zero-point energy and thermal vibrational contributions at 300~K amount to about $0.1$~eV, favoring the low-spin state,\cite{Kepp2016,Pierloot2006-iq}
which shifts the target range of the adiabatic energy difference to roughly $0.1$--$0.4$~eV.
Moreover, semi-local density functionals such as PBE are known to systematically overestimate spin-state energy splittings by several tenths of an eV, motivating the use of a broader energy window extending up to $\sim 1$~eV when screening candidate spin-crossover materials \cite{KEPP2013196,Kepp2016,Vela2020,Mariano2020-nx,Mariano2021-wu,C3CP55506B}.

\paragraph{Fixed-geometry approximation and label noise}
The $\Delta E_\text{H--L}$ criterion of (0,1) eV applies when the geometries for both spin states are fully optimized; however, in this work, $\Delta E_\text{H--L}$ is evaluated using geometries from the QMOF dataset\cite{ROSEN20211578}. These are optimized using spin-polarization using PBE+D3(BJ), as we do here, and by initializing magnetic moments on d- and f-block elements and allowing them to relax freely to a self-consistent local minimum. 
The adiabatic energy differences computed within this fixed-geometry approximation may differ to a non-negligible extent from those obtained after full spin-state-specific structural optimization, thereby introducing noise into the dataset.

\begin{figure}[t]
\includegraphics[scale=0.8,
trim=9cm 0cm 9cm 0cm,
clip]{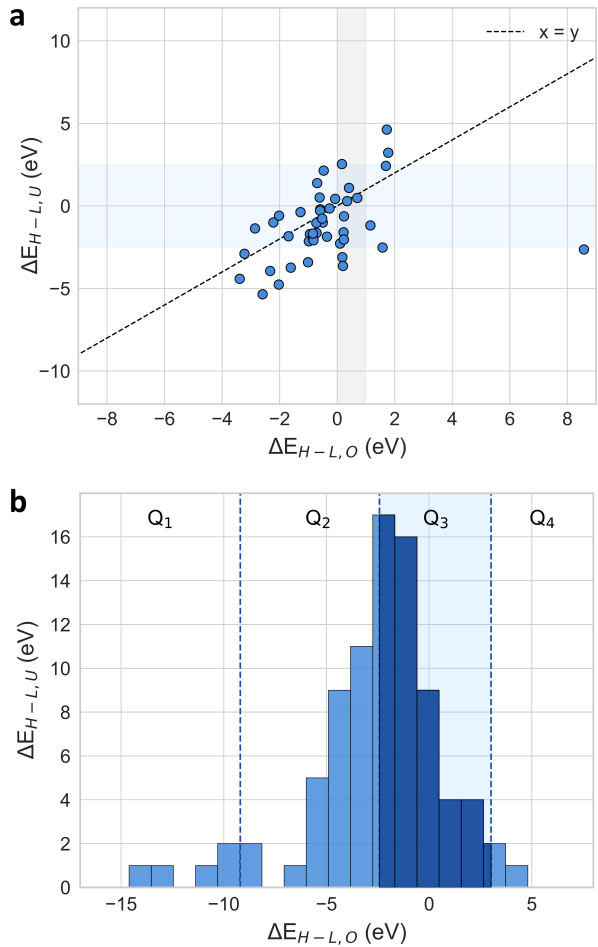}
\caption{(a) Plot showing $\Delta E_{H-L,O}$, obtained using the \texttt{SCO-MOF-RelaxWorkChain} vs $\Delta E_\text{H--L,U}$, obtained using the \texttt{SCO-MOF-SCF-WorkChain} using unoptimized structures. The grey shaded region corresponds to an estimate of the values of $\Delta E_\text{H--L,O}$ that are potentially interesting for SCO, and the corresponding region for $\Delta E_\text{H-L,U}$ is shaded in blue. The diagonal indicates the $x=y$ line.
(b) Distribution of the labels $\Delta E_\text{H--L,U}$ of the test set. The range of the labels is partitioned into four quantiles, Q$_1$, Q$_2$, Q$_3$ and Q$_4$, with Q$_3$ being the quantile of interest based on (a).}
\label{fig:5}
\end{figure}

To identify the appropriate quantile range, the relationship between $\Delta E_\text{H--L}$ values computed with and without relaxation at the respective spin states is analyzed, and shown in Figure \ref{fig:5}(a). Values computed for the subset of test set for which both spin state calculations converged, are used to determine the quantiles.
In Figure \ref{fig:5}(a), $\Delta E_\text{H--L,O}$ refers to values computed using geometries relaxed at the corresponding spin state, while $\Delta E_\text{H--L,U}$ refers to values obtained using the same geometry for both spin states, which is taken from the QMOF database. The subscripts O and U denote "optimized" and "unoptimized", respectively. The plot shows a substantial correlation. Notably, the true range of interest, 0 to 1 eV in $\Delta E_\text{H--L,O}$ approximately maps to -2.5 to 2.5 eV in $\Delta E_\text{H--L,U}$. Consequently, this broader range is adopted as the quantile of interest for QRT-AL, and $\Delta E_\text{H--L,U}$ is used as the label. It is worth noting that unoptimized values can be interpreted as the noisy counterpart of optimized implying that our data is not only limited but also noisy.

\paragraph{Quantile definition and weights in QRT-AL}
Figure \ref{fig:5} (b) illustrates the distribution of $\Delta E_\text{H--L,U}$ for the test set, with the shaded region highlighting the quantile of interest determined, Q$_3$. The remainder of the label space is segmented into three quantiles, Q$_1$, Q$_2$ and Q$_4$, also shown in the figure. The quantiles Q$_1$, Q$_2$, Q$_3$ and Q$_4$, based on the labels of the test set, correspond to the ranges (0, 0.05), (0.05, 0.40), (0.40, 0.95), and (0.95, 1.00), respectively. Their associated weights are 0.05, 0.2, 0.7, and 0.05. The highest weight is assigned to the quantile of interest, and the weights assigned to other quantiles were determined based on the distribution of the labels, with progressively lower weights assigned to regions farther from Q$_3$. They are kept non-zero, however, to ensure that the training set also includes some samples from those regions, improving model generalization, as proposed in Ref. \citenum{qrtal}.

\begin{figure*}[ht]
\centering
\includegraphics[scale=0.67,
trim=0cm 4.5cm 1.7cm 4cm,
clip]{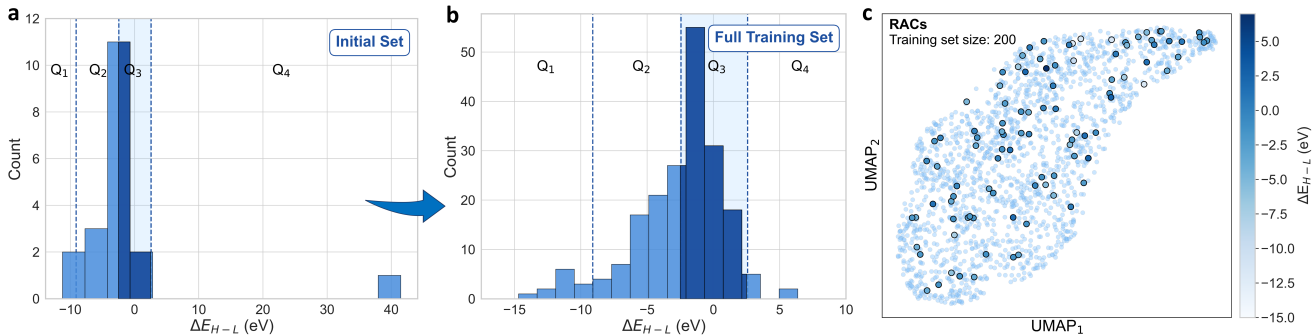}
\caption{(a) Histogram depicting the $\Delta E_\text{H--L}$ distribution of the 20 initial MOFs selected randomly, and (b) distribution of the full training set of 200 MOFs, sampled using QRT-AL iteratively. The vertical lines indicate the different quantiles used in QRT-AL, and the light blue shaded region shows the quantile of interest.
(c) UMAP of the training pool consisting of 1862 MOFs is presented in the RACs feature space. The MOFs selected using QRT-AL are highlighted using darker markers, and are colored by their $\Delta E_\text{H--L}$ values computed in this work using DFT.}
\label{fig:6}
\end{figure*}

\subsection{Descriptors}
The set of descriptors used in this work to predict $\Delta E_\text{H--L}$ is Revised Auto-Correlations (RACs \cite{racs}). RACs comprise products and differences of heuristic atomic properties on graphs of the atomic structures. Formally, a typical product-based RAC is computed on the graph representation of the structure as follows:

\begin{equation}
    P_d^{prod} = \sum_{i}^{\text{start}} \sum_{j}^{\text{scope}}  P_iP_j \delta (d_{ij},d).
\end{equation}

Two atom lists, called \textit{start} and \textit{scope}, are required to compute RACs. The atomic property $P$ of atom $i$ selected from the \textit{start} atom list is correlated to atom $j$ from the \textit{scope} atom list, when they are separated by $d$ number of bonds. Further details on RACs can be found in Supplementary Section S1.

This descriptor was chosen as it has proven to be meaningful in establishing structure-property relationships in transition metal complexes \cite{racs, doi:10.1021/acs.iecr.8b04015} and metal-organic frameworks \cite{racs_mofs}; capturing both metal and ligand chemistry, and predicting spin splitting energies \cite{racs} as well as SCO transition temperatures \cite{doi:10.1021/acs.jpca.3c07104}.
These descriptors were generated using molSimplify \cite{molSimplify}. The features include contributions centered on metals, linkers, and functional groups, and are weighted by atomic properties including atom identity, connectivity, Pauling electronegativity, covalent radii, nuclear charge, and polarizability. Averaging over all atoms in each MOF yielded a set of 156 features. This feature-set is employed to select samples using QRT-AL, as well as to train ensemble tree-based models to predict $\Delta E_\text{H--L}$. 

\subsection{Training set construction}

RACs were computed for the remaining 2084 MOFs, and 1862 of them could be featurized using molSimplify. This set of 1862 MOFs constitutes the training pool, from which the training set is sampled using QRT-AL. The first 20 MOFs were selected using random sampling and the corresponding $\Delta E_\text{H--L,U}$  was computed using the AiiDA workflow ($U$ dropped from the subscript hereafter). These labels, together with the features, were used to construct the initial regression tree, for which the minimum samples in each leaf was set to 5, as suggested in prior works \cite{ashna, jacs_ashna, qrtal}. QRT-AL was subsequently used to determine the number of MOFs to be labeled from each leaf, after which they were labeled using the AiiDA workflow. At each active learning iteration, the threshold values corresponding to the predefined quantiles were recalculated based on the true labels available in that round. This process continued until a total of 200 MOFs (i.e. approximately 10\% of the MOF-2184 dataset) were labeled, with 20 MOFs being labeled in each iteration, followed by retraining of the regression tree. When the workflow failed for certain MOFs (i.e., the SCF calculations did not converge), new MOFs were resampled from the same leaves. The combined dataset of 276 MOFs, for which $\Delta E_\text{H--L}$ was computed using DFT, is referred to as the computed-SCO-276 (cSCO-276) dataset. 

Figure \ref{fig:6}(a) shows the distribution of the labels of the initial training set sampled randomly and Figure \ref{fig:6}(b) shows the corresponding distribution for the final training set (see S3 for the distribution of the labels in each active learning round). QRT-AL successfully sampled a high proportion of MOFs within the noisy quantile of interest, i.e., MOFs with $\Delta E_\text{H--L}$ values between -2.5 and 2.5 eV. As intended, fewer MOFs were sampled from regions of the label space outside this quantile, reflecting the assigned weights. Although the quantiles were estimated based on a small test set (fewer than 100 data points), QRT-AL successfully selected the relevant MOFs using only a minimal number of labeled samples. It is noteworthy that the MOFs selected in the training set are also representative and diverse in the feature space as shown in the UMAP of the training pool in the RACs feature space (Figure \ref{fig:6}(c)). This set of 200 MOFs therefore constitutes an informative training set that can be leveraged for predicting adiabatic energy differences and identifying potential candidate MOFs that exhibit SCO.

\subsection{Model training and testing}

\begin{figure*}[ht]
\centering
\includegraphics[scale=0.63,
trim=0.1cm 4cm 0.2cm 4cm,
clip]{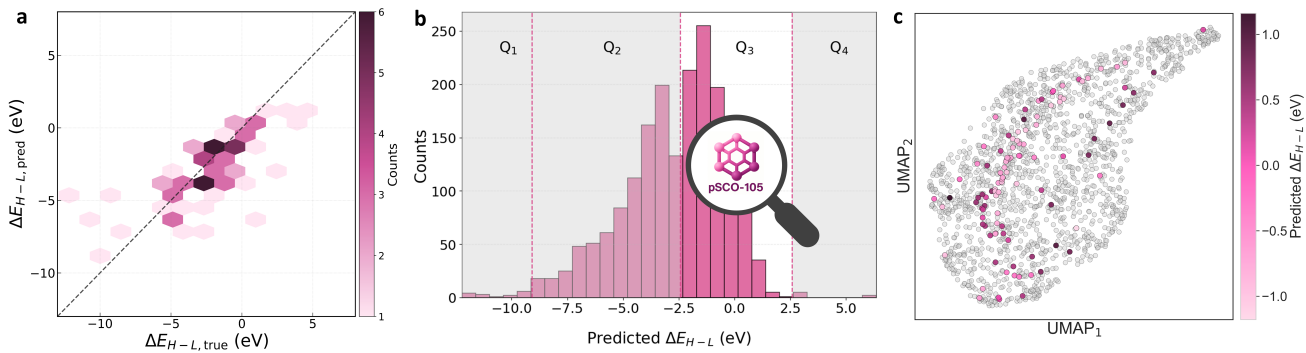}
\caption{
(a) Parity plot using hex-bins of the true vs predicted values of $\Delta E_\text{H--L}$ of the test set, using the RF (RACs) model. 
(b) Distribution of the predicted values of $\Delta E_\text{H--L}$ of the remaining 1662 unlabeled MOFs in the training pool, obtained using the RF (RACs) model. The vertical lines show the different quantiles used in QRT-AL, the quantile of interest is highlighted. The high confidence pSCO-105 set is obtained using uncertainty predictions from quantile random forests.
(c) UMAP of the total training pool consisting of 1862 MOFs is presented in the RACs feature space. The MOFs with high confidence $\Delta E_\text{H--L}$ predictions in the range of interest are shown in shades of pink.
}
\label{fig:7}
\end{figure*}

Random Forest (RF \cite{rf}) models were trained on the dataset curated by QRT-AL. Tree-based models were selected as the final predictors due to their compatibility with the decision-tree principles underlying QRT-AL and their demonstrated efficiency in low-data regimes for tabular datasets \cite{rf_vs_gcn, NEURIPS2022_0378c769}. Three distinct models were developed using different feature representations: RACs, ST-120 descriptors, and their concatenation (RACs + ST-120). Features were computed for both the training and test sets; however, 11 of the 87 MOFs in the test set could not be featurized using RACs, reducing the test set size to 76. To ensure robustness, three outliers (one from the training set and two from the test set) were removed, resulting in final set sizes of 199 and 74 MOFs, respectively. \texttt{GridSearchCV} in Scikit-learn \cite{scikit-learn} was used to determine the best hyperparameters of each model (min-samples-leaf, n-estimators, and max-features). For comparative purposes, a graph-based Crystal Graph Convolutional Neural Network (CGCNN \cite{cgcnn}) model, with 100 training epochs and a hidden dimension of 64, was also trained. CGCNNs are commonly used for MOF property prediction tasks \cite{ROSEN20211578} and thus serve as a relevant baseline.

The parity plot in Figure \ref{fig:7}(a) compares the true and predicted $\Delta E_\text{H--L}$ values for the test set. Model performance was evaluated using two metrics: the Mean Absolute Error (MAE) across the full test set, and the Quantile Mean Absolute Error (QMAE), which restricts the MAE calculation to the quantile of interest (-2.5 to 2.5 eV). As shown in Table \ref{table2}, the RF model trained with RAC descriptors achieved the lowest errors for both metrics (MAE = 1.488 eV, QMAE = 1.218 eV), confirming that RACs are the most informative features for predicting spin-splitting energetics in MOFs. The lower QMAE relative to MAE highlights the effectiveness of QRT-AL in focusing predictive accuracy on the targeted $\Delta E_\text{H--L}$ range, even with limited training data. This intentional imbalance in the training set selection explains the higher MAE values, as the model prioritizes accuracy in the SCO-relevant region over the entire output space.

\renewcommand{\arraystretch}{1.18}
\begin{table*}[ht]
\centering
\caption{Performance metrics of different ML models for predicting $\Delta E_\text{H--L}$ values of the test set. The table compares the mean absolute error (MAE), quantile mean absolute error (QMAE), the \% of MOFs in the optimized test subset whose true $\Delta E_\text{H--L}$ falls in the quantile of interest (0-1 eV) and is correctly predicted in the approximate quantile of interest (-2.5, 2.5) eV ($\%\text{ Recovered}$ or recall value), balanced accuracy, and the confusion matrix components: true positives (TP), true negatives (TN), false positives (FP), and false negatives (FN). TP, TN, FP, and FN values correspond to the binary evaluation of MOFs into the quantile of interest (positive class) versus all others (negative class). The best-performing method and statistically equivalent results (t-test at the 0.05 level), are highlighted in bold and italics, respectively, for the first four performance metrics.}

\label{table2}
\begin{tabular}{lcccccccc}
\hline
Method & MAE (eV) & QMAE (eV) & \% Recovered & Balanced Acc. & TP & TN & FP & FN \\
\hline
\textbf{RF (RACs)}          & \textit{1.488} & \textbf{1.218} & \textbf{81.82} & \textbf{72.58} & 9 & 19 & 11 & 2 \\
RF (ST-120)        & 2.289 & 1.672 & 63.64 & 60.15 & 7 & 17 & 13 & 4 \\
RF (RACs + ST-120) & 1.584 & 1.335 & \textit{81.82} & \textit{72.58} & 9 & 19 & 11 & 2 \\
CGCNN              & \textbf{1.464} &  \textit{1.223} & \textit{81.82} & 70.91 & 9 & 18 & 12 & 2 \\
\hline
\end{tabular}
\end{table*}

Given the noise in the $\Delta E_\text{H--L}$ labels, and to further assess the model’s ability to identify SCO-active MOFs, we reformulated the prediction task as a binary evaluation: MOFs with predicted $\Delta E_\text{H--L}$ values falling within the targeted quantile were considered "SCO-active candidates" (positives), and thus potentially capable of exhibiting spin crossover near room temperature, while all others were treated as negatives. 
We used a recall-based assessment (termed \% Recovered), measuring the proportion of MOFs with accurately computed $\Delta E_\text{H--L}$ values (from fully optimized geometries) that fall within the SCO-relevant window (0-1 eV) and are correctly predicted by the model to lie in the broader range of (-2.5, 2.5) eV. This broader range accounts for the prediction uncertainty while ensuring high-quality reference data for validation.
This approach leverages the regression model’s predictions to evaluate its sensitivity and specificity in detecting potential SCO materials. The subset of 41 MOFs from the test set with $\Delta E_\text{H--L}$ values computed from fully optimized geometries were used for this analysis, and the results are summarized in Table \ref{table2}.

For the RF (RACs) model, the high \% Recovered (81.8\%)  demonstrates the model’s effectiveness in identifying SCO-active candidates, despite the limited training set size (199 samples). The model also attains a high {\sl balanced accuracy} of 72.6\%, defined as the average of the proportion of correctly classified samples in each class, which is particularly appropriate for evaluating performance on imbalanced datasets such as ours.
Table \ref{table2} also reports the numbers of true positives (TP), true negatives (TN), false positives (FP), and false negatives (FN). The confusion matrix reveals a low false-negative rate (FN=2), indicating that the model rarely misses true SCO-active MOFs: a critical advantage for materials discovery. The false-positive rate (FP=11) is moderate but acceptable, as false positives can be efficiently verified through follow-up DFT calculations.
In contrast, the RF model trained on ST-120 descriptors performs significantly worse, with a \% Recovered of 63.6\% and a balanced accuracy of 60.2\%. This underscores the limited predictive power of ST-120 features for this task. Interestingly, combining RACs and ST-120 descriptors does not improve performance, suggesting that the additional information from ST-120 does not enhance the model’s ability to capture the relevant structure-property relationships.

The CGCNN model achieved a QMAE of 1.223 eV, comparable to the RF model with RAC descriptors. Moreover, its \% Recovered (81.8\%) and balanced accuracy (70.9\%) was slightly lower (or equal).
These results suggest that, for small datasets, classical machine learning approaches (e.g., RF) can outperform or match the performance of graph-based neural networks, likely due to the latter’s higher data requirements for effective training. The similarity in QMAE but lower balanced accuracy of CGCNN also highlights the robustness of RF (RACs) in handling class imbalance, a critical factor in active learning scenarios.

\subsection{Predicting spin-crossover behavior}

\paragraph{Model generalization beyond the training distribution}
To further evaluate our approach, we assessed our model on a total of 9 systems comprising reference molecules, transition-metal complexes, and MOFs reported in the literature. Given its superior predictive performance, the RF (RACs) model was employed to predict $\Delta E_\text{H--L}$ for these systems. The MOF with QMOF-ID \textit{qmof-a5e17b9} (CSD refcode: LOJLAZ) is a known spin-crossover MOF \cite{Clements2014, ROSEN20211578}, but it is absent from our dataset of 2184 MOFs since it contains two metal types. This system thus represents an out-of-distribution case for our task since the models in this work were trained exclusively with MOFs with a unique metal type. Despite this, the model predicts a $\Delta E_\text{H--L}$ of -1.08 eV for this MOF, which falls within the relevant range of interest for SCO activity, demonstrating that our model correctly classifies this MOF as SCO-active. In addition, we evaluated the model against reference data for known transition-metal complexes\cite{hoffman_exp,Hoffman_theo,XANLEE_exp,Mariano2023,phen_exp,Vela2020,Mariano2021-wu,bpy_theory,abpt_exp,tacn_theory} (see Table S1 in the SI). In all cases, the model correctly identifies the positives, predicting $\Delta E_\text{H--L}$ values within the SCO-relevant range. Notably, for the Fe$_2$(tpt)$_2$(NCS)$_4$\cite{XANLEE_exp,Mariano2023} system, two cases were tested: with and without H$_2$ guest molecules. Interestingly, the model predicts a $\Delta E_\text{H--L}$ of -2.4 eV for the bare material (close to the threshold for a high spin ground state) and -0.02 eV for the adsorbed H$_2$, indicating a SCO-consistent behavior in the presence of the gas. This is consistent with previous theoretical studies \cite{Mariano2023} and experiments \cite{XANLEE_exp} showing that the low spin state can be stabilized by the strong interaction with the guest molecule. Overall, these results show that, despite being trained on a limited and noisy dataset, the model captures chemically meaningful trends across a diverse set of SCO-active systems.

\paragraph{Prediction with high-confidence in pSCO-105}
The RF (RACs) model was used to predict $\Delta E_\text{H--L}$ across the remaining 1662 MOFs in the training pool, aiming to identify SCO-active candidates. As shown in Figure \ref{fig:7}(b), 843 MOFs were predicted to fall within the SCO-relevant window ($-2.5$ to $2.5$ eV), significantly narrowing the search space. 

To further isolate high-confidence candidates, prediction uncertainty was quantified using a Quantile Random Forest (QRF \cite{Meinshausen2006QuantileRF}). Unlike standard random forests, which predict only the mean $\Delta E_\text{H--L}$, QRF estimates conditional quantiles, here the 5th (Q$_{0.05}$) and 95th (Q$_{0.95}$) percentiles. A MOF was classified as a high-confidence SCO candidate if both quantiles lay within the target window, i.e.
\[
-2.5 < Q_{0.05} < Q_{0.95} < 2.5 .
\]
This criterion ensures 95\% confidence in the SCO-active prediction. Using the \texttt{RandomForestQuantileRegressor} implementation from Scikit-Learn, quantile predictions were generated for the training pool. Out of the 843 MOFs with $\Delta E_\text{H--L}$ values in the (-2.5, 2.5) eV range, 105 MOFs met the high-confidence criterion. 
Figure \ref{fig:7}(c) shows the UMAP of the total training pool of 1862 MOFs in the RACs feature space, with the subset of 105 MOFs with high-confidence predictions within the target range highlighted in shades of pink. This dataset, named the predicted-SCO-105 (pSCO-105) represents a set of MOFs with a high predicted likelihood of spin-crossover behavior.

Surprisingly, cobalt-based MOFs dominate the pSCO-105 set, with only two iron-based and one nickel-based structures included. Notably, despite similar representation of cobalt and manganese in the training pool (709 Co-based vs 501 Mn-based MOFs), no Mn-containing MOFs met the high-confidence threshold. 
Regarding coordination environments, consistent with known trends in SCO chemistry\cite{sco-review1,sco-review2}, 103 structures exhibit octahedral coordination, while only one structure each shows coordination numbers of four and five (qmof-b808abd, Ni; qmof-472da64, Co).
%This is potentially due to the high number of 6-fold coordinated MOFs present in the training set, and the MOF-2184 database in general.

\section{Discussion}
We developed an end-to-end workflow to predict $\Delta E_{\text{H--L}}$ and identify spin-crossover (SCO) MOFs. Starting from the QMOF database, we curated a subset of 2184 candidates MOFs potentially exhibiting SCO behavior. A tree-based active learning strategy, Quantile RT-AL, was employed to select informative training samples from this subset, with a specific focus on $\Delta E_{\text{H--L}}$ values within the region of interest. This region was defined based on target $\Delta E_{\text{H--L}}$ values obtained from self-consistent field calculations at the low- and high-spin states using the PBE+D3(BJ) functional. Using QRT-AL in combination with the SCO-MOF workflow implemented within the AiiDA framework, $\Delta E_{\text{H--L}}$ values were iteratively computed for a total of 200 MOFs. This process yielded the \textbf{computed SCO (cSCO) database}, a curated collection of 276 MOFs from both the test and training sets, characterized by $\Delta E_{\text{H--L}}$ values obtained via single-point DFT calculations (cSCO-276).

The labeled dataset was used to train ensemble tree-based random forest (RF) regression models. The RF model trained using revised autocorrelation (RAC) descriptors achieved the lowest MAE and QMAE when compared to ST-120 descriptors, indicating that RACs provide the most informative representation for predicting spin-splitting energies in MOFs. The RF (RACs) model was also compared against a crystal graph convolutional neural network (CGCNN) model, and the performances were similar, showing that there is no benefit to training deep learning models in the low-data regime. To assess the model’s ability to identify spin-crossover behavior, we further performed a binary evaluation. This evaluation was conducted using a subset of 41 MOFs for which geometry-optimized $\Delta E_{\text{H--L,O}}$ values were available. 

The RAC-based RF model achieved a recall of 82\% (i.e., it correctly identified 82\% of the SCO-active MOFs) along with a high balanced accuracy of 73\%, demonstrating strong SCO-identification performance despite the small training set ($n=199$). This result highlights the effectiveness of QRT-AL in constructing models that are both accurate and robust to uncertainty. Analysis of the confusion matrix further illustrates the reliability of the RAC-based RF model, which correctly identifies most true positives (9) and true negatives (19), while producing very few false negatives (2). This high sensitivity is particularly important for SCO discovery, as it minimizes the risk of overlooking viable candidates. The moderate number of false positives (11) reflects the intentionally imbalanced training-set selection induced by QRT-AL and is acceptable in a screening context, where false positives can be filtered in subsequent validation steps. Finally, the model also successfully identified known SCO-active molecules, complexes, and MOFs outside the training distribution, demonstrating its ability to generalize beyond the curated dataset. Overall, these results show that RAC-based RF models trained on a small, actively selected dataset effectively capture the structure-property relationships required to identify SCO-active MOFs. 

The RF (RACs) model was used to make predictions on the remaining training pool, and 843 MOFs fell within the SCO-active range of interest. Uncertainty quantification was then performed using Quantile Random Forests (QRF) to further filter this set. 105 MOFs were found to have $\Delta E_{\text{H--L}}$ values in the SCO-active range with 95\% confidence. This set constitutes the predicted SCO (pSCO-105) database, comprising MOFs with a high likelihood of exhibiting spin-crossover behavior.

Our computational screening of the MOF-2184 dataset operates within a regime of inherent numerical and physical uncertainty that affects the precision of the calculated $\Delta E_{\text{H--L}}$ values. Primary sources of this label noise include the fixed-geometry approximation—where the use of unrelaxed structures introduces a significant offset relative to the true adiabatic energy difference—and the known systematic biases of the PBE+D3(BJ) functional. %Furthermore, while the spin-crossover phenomenon technically involves the transition between the ground state and the first metastable spin state, our current workflow simplifies this by considering only the lowest and highest spin configurations; a more granular treatment of intermediate spin states remains a subject for future work.
Furthermore, while spin crossover is an interconversion between spin states of different multiplicity that are closest in Gibbs free energy (and, as such, in practice typically involves the ground state and the lowest-lying state of a different multiplicity), here we restrict the workflow to the lowest- and highest-spin configurations. Intermediate-spin states, which are nonetheless usually higher in energy for predominantly octahedral environments, and an overall unbiased exploration of the spin energy-landscape, are deferred to future work.

Despite these cumulative approximations, the Quantile Active Learning (QRT-AL) framework demonstrates remarkable robustness in identifying the underlying SCO-active manifold. Crucially, the utility of this approach remains independent of the specific level of theory employed. Even if an idealized, perfectly accurate method for computing $\Delta E_{\text{H--L}}$ were to become available, the active learning strategy would remain an essential tool for navigating the vast, high-dimensional chemical space of MOFs where exhaustive electronic-structure labeling remains computationally prohibitive.

As a perspective, MOFs from pSCO-105 can be screened for various applications, such as spin-crossover assisted gas capture and release \cite{Mariano2023}. Future work may also benefit from more accurate exchange-correlation treatments, including r2SCAN\cite{r2scan} or machine-learned functionals designed for accurate predictions of $\Delta E_{\text{H--L}}$ \cite{De_Mendonca2023}. Additionally, the Quantile RT-AL approach could be extended to other rare materials phenomena, where goal-oriented data acquisition may be necessary. Overall, this study demonstrates that data-efficient machine-learning workflows can reliably capture complex phenomena such as spin crossover using limited and noisy data. The success with unoptimized $\Delta E_{\text{H--L}}$ values suggests a practical route for rapid pre-screening for SCO MOFs before refined, expensive experimental or computational investigations.

\section{Methods}

\subsection{Electronic structure calculations and workflow}
Two automated workflows were developed using the AiiDA workflow manager \cite{PIZZI2016218, aiida, martin_uhrin}. 
AiiDA, implemented in Python, includes a \textsc{Quantum ESPRESSO} plugin \cite{Giannozzi2009}, which is used here to run DFT calculations.
The two AiiDA workflows developed around this plugin are named \texttt{SCO-MOF-RelaxWorkChain} and \texttt{SCO-MOF-SCF-WorkChain}. Version 7.1 of \textsc{Quantum ESPRESSO} was employed. 
In the \texttt{SCO-MOF-RelaxWorkChain}, the geometry is first optimized at each spin state. A final SCF calculation is then performed using the relaxed geometries, with the \texttt{PwBaseWorkChain}.
The \textit{total magnetization} of the system is kept fixed. This approach was chosen after initially testing the input parameter \textit{starting magnetization} in \textsc{Quantum ESPRESSO}, which typically led the system to converge to its ground state spin rather than the targeted spin state. Thus, for each MOF, \textit{total magnetization} was provided as an input, which was computed using the oxidation states as described previously (Table \ref{tab:oximachine}). 
The total magnetization for systems with multiple transition metals (of the same type) was obtained by simply multiplying the value of one by the number of metals in the system. PseudoDojo pseudo potentials \cite{VANSETTEN201839} were used via the AiiDA pseudo plugin, and the wave-function and charge density cut-offs recommended by AiiDA (exclusively for each case) were used for each system.  PBE\cite{PhysRevLett.77.3865} was used for the exchange and correlation together with the Grimme-D3 correction \cite{Grimme2010} and the BJ damping scheme \cite{Smith2016RevisedDP} for long-range interactions.  Note that the r2SCAN functional was initially tested in this pipeline, however it was eventually abandoned due to severe convergence difficulties. The BFGS algorithm was used for relaxation until the residual forces acting on atoms were less than 0.001 eV/\AA{}, and the convergence threshold for the total energy was 0.0001 Ryd. The convergence criterion for total energy in the SCF calculations was set at $10^{-6}$ Ryd. The Brillouin zone is sampled by setting the \textit{kpoints-distance} parameter to 0.5/\AA{}, such that the grid adapts based on the system. Given the difficulty in achieving convergence to non-ground-state spin configurations, the BFGS optimizer was allowed a maximum of 100 steps, while the SCF cycle was permitted up to 400 iterations to improve the likelihood of successful convergence. 

The \texttt{SCO-MOF-SCF-WorkChain} is a simplified version of the former that omits structural relaxation. It is intended either for computing LS and HS energies on the same geometry or when optimized LS and HS structures are available and SCF calculations are to be performed using a different set of parameters.

\section{Associated Content}

\subsection{Data availability statement}
The Python code implementing the Quantile Regression Tree-based Active Learning algorithm has been made publicly available on GitHub at \href{https://github.com/AshnaJose/SCO-MOF-Design}{https://github.com/AshnaJose\\/SCO-MOF-Design}. The repository includes a comprehensive example of using QRT-AL with a MOF dataset. The repository also contains the AiiDA Quantum ESPRESSO workflows (\texttt{SCO-MOF-RelaxWorkChain} and \texttt{SCO-MOF-SCF-WorkChain}) developed in this work. The different descriptors computed for model training and the DFT-computed $\Delta E_\text{H--L}$ values for the training and test set (cSCO-276 dataset) are also provided in the repository. The high-confidence predicted $\Delta E_\text{H--L}$ values obtained using the QRF (RACs) model for the pSCO-105 subset are also provided.

\subsection{Supporting Information}
The Supporting Information is available free of charge at SX. It includes descriptions of the descriptors employed for model training (ST-120 and RACs), the details of the AiiDA workflows (\texttt{SCO-MOF-RelaxWorkChain} and \texttt{SCO-MOF-SCF-WorkChain}) developed for DFT calculations of spin-state energetics, and the specific computational settings used. Additionally, it provides label distributions of $\Delta E_\text{H--L}$ across successive rounds of Quantile Regression Tree-based Active Learning (QRT-AL). Finally, RF (RACs) model predictions for out-of-distribution known spin-crossover complexes and MOFs are also provided.

\begin{acknowledgement}
We acknowledge CINES, IDRIS and TGCC under project No.~A0200907211 and A0180907211, INP2227/72914/gen5054, 
as well as CIMENT/GRICAD for computational resources. This work has been partially supported by MIAI Cluster (ANR-23-IACL-0006). Discussions within the French collaborative network in artificial intelligence in materials science GDR CNRS 2123 (IAMAT) are also acknowledged. This work was partially supported by the French government “France 2030” initiative, under the DIADEM program managed by the “Agence Nationale de la Recherche” (ANR-22-PEXD-0015, DIAMOND). This work has benefited from a French government grant managed by the Agence Nationale de la Recherche under the France 2030 program, reference ANR-23-IACL-0006.
\end{acknowledgement}

\bibliography{references}
\end{document}